\documentclass[12pt,preprint]{aastex}

\newcommand{\oiii}{[\ion{O}{3}]}
\newcommand{\niii}{[\ion{N}{3}]}
\newcommand{\oi}{[\ion{O}{1}]}
\newcommand{\nii}{[\ion{N}{2}]}
\newcommand{\cii}{[\ion{C}{2}]}
\newcommand{\siii}{[\ion{Si}{2}]}
\newcommand{\ci}{[\ion{C}{1}]}
\newcommand{\um}{\,$\mu$m}

\begin{document}


\title{Large Silicon Abundance in Photodissociation Regions\footnote{Based on observations with ISO, an ESA project with instruments funded by ESA Member States (especially the PI countries: France, Germany, the Netherlands and the UK) and with the participation of ISAS and NASA.}}

\author{Yoko Okada and Takashi Onaka}
\affil{Department of Astronomy, Graduate School of Science, University of Tokyo, Bunkyo-ku, Tokyo, 113-0033, Japan}
\email{okada@astron.s.u-tokyo.ac.jp; onaka@astron.s.u-tokyo.ac.jp}

\author{Takao Nakagawa}
\affil{Department of Infrared Astrophysics, Institute of Space and Astronautical Science (ISAS), Japan Aerospace Exploration Agency (JAXA), Kanagawa, 229-8510, Japan}

\author{Hiroshi Shibai}
\affil{Graduate School of Science, Nagoya University, Nagoya, 464-8602, Japan}

\author{Daigo Tomono}
\affil{Subaru Telescope, National Astronomical Observatory of Japan, Hilo, Hawaii 96720, U.S.A.}

\and

\author{Yukari Y. Yui}
\affil{Earth Observation Research and Application Center, Office of Space Applications, Japan Aerospace Exploration Agency (JAXA), Tokyo, 104-6023, Japan}

\begin{abstract}
We have made one-dimensional raster-scan observations of the $\rho$ Oph and $\sigma$ Sco star-forming regions with two spectrometers (SWS and LWS) on board the {\it ISO}.  In the $\rho$ Oph region, \siii~35\um, \oi~63\um, 146\um, \cii~158\um, and the H$_2$ pure rotational transition lines S(0) to S(3) are detected, and the PDR properties are derived as the radiation field scaled by the solar neighborhood value $G_0\sim 30$--500, the gas density $n\sim 250$--$2500$~cm$^{-3}$, and the surface temperature $T\sim 100$--$400$~K.  The ratio of \siii~35\um~to \oi~146\um~indicates that silicon of 10--20\% of the solar abundance must be in the gaseous form in the photodissociation region (PDR), suggesting that efficient dust destruction is undergoing even in the PDR and that part of silicon atoms may be contained in volatile forms in dust grains.  The \oi~63\um~and \cii~158\um~emissions are too weak relative to \oi~146\um~to be accounted for by standard PDR models.  We propose a simple model, in which overlapping PDR clouds along the line of sight absorb the \oi~63\um~and \cii~158\um~emissions, and show that the proposed model reproduces the observed line intensities fairly well.  In the $\sigma$ Sco region, we have detected 3 fine-structure lines, \oi~63\um, \nii~122\um, and \cii~158\um, and derived that 30--80\% of the \cii~emission~comes from the ionized gas.  The upper limit of the \siii~35\um~is compatible with the solar abundance relative to nitrogen and no useful constraint on the gaseous Si is obtained for the $\sigma$ Sco region.
\end{abstract}



\section{Introduction}

Silicon and carbon are major constituents of interstellar grains \citep{Mathis}.  The gas phase Si abundance is about 5\% of solar in cool clouds \citep{SavageSembach} and the depleted atoms are thought to reside in interstellar grains.  According to the observations of UV absorption lines, Si shows a systematic trend of depletion with the density of the diffuse cloud as well as Mg and Fe \citep{Jenkins,Harris}, whereas the depletion of Fe is suggested not to continue to increase with either density or extinction for $\langle n_\mathrm{H} \rangle \gtrsim 1$~cm$^{-3}$ \citep{Snow}.  \citet{Sofia94} suggested that Fe has the greatest fraction of its atoms incorporated into dust followed by Mg and Si, and suggested oxides and/or metallic Fe as the grain core population.  \citet{Fitzpatrick} showed that the gas-phase abundance of Si and Fe is well correlated with each other to some extent, but the degree of returning to the gas phase is different between the two: it is easier for Si to return to the gas than for Fe.  This suggests that at least part of Si and Fe atoms constitute independent grain populations.  \citet{Jones00} examined the relation among the depletion of Si, Mg, and Fe, suggesting that Si in dust is preferentially eroded with respect to Mg and that both of these elements are preferentially eroded with respect to Fe.  This cannot completely be explained by mechanical sputtering process, indicating some chemically selective processes involved or the presence of Si in phases other than refractory silicates.  \citet{Frisch} showed that the gas-to-dust ratio in the local ISM is well correlated with the fraction of the dust mass carried by Fe.  This indicates that Fe forms a robust core that is not destroyed during grain processing in the ISM.  \citet{Cartledge04} showed that the dust phase abundance of O derived with the ``missing mass'' method is consistent with the grain model in which silicates are primarily Mg based and most or all of Fe is in metal or oxides.  On the other hand, C, N, and O abundance is relatively constant against the gas density \citep{Sofia97,Cardelli,Andre,Jensen}, although the dependence on the density for oxygen is also reported \citep{Cartledge01,Cartledge04}.

Depletion in active star-forming regions, where the density is much higher than diffuse clouds, has been studied by infrared line emissions.  Intense \siii~35\um~emission has been reported in a shocked region in Orion \citep{Haas}, the Galactic center region \citep{Stolovy}, the Carina region \citep{Mizutani}, and the starburst galaxy NGC~253 \citep{Carral}.  Not only in ionized regions but also in photodissociation regions (PDRs) large gas phase abundances of Si have been reported: 30\% of the solar abundance in Sharpless 171 (S171) \citep[][hereafter OOSD03]{Okada}; 50\% in G333.6-0.2 \ion{H}{2} region \citep{Colgan}; 20--30\% in the reflection nebula NGC~7023 \citep{Fuente00}; 10\% in OMC-1 \citep{Rosenthal}.  \citet{YoungOwl} observed nine reflection nebulae and detected \siii~35\um~in two regions with the density range of $0.9$--$2\times 10^4$~cm$^{-3}$ and indicated that the intensities of \siii~35\um~in those regions agree with the PDR model.  The conversion of the \siii~35\um~intensity to the Si abundance depends on the physical properties, especially the gas density, and thus we need to derive those properties from other line emissions.

In this paper, we report detection of intense \siii~35\um~emission in the $\rho$ Oph reflection nebula based on observations with the Short-Wavelength Spectrometer \citep[SWS;][]{deGraauw} and the Long-Wavelength Spectrometer \citep[LWS;][]{Clegg} on board the {\it Infrared Space Observatory} \citep[{\it ISO};][]{Kessler}.  We discuss the constraint on the Si abundance in the PDR gas, as well as the physical properties of the PDR.  Preliminary results have been reported in \citet{Tomono}.  The $\rho$ Oph region is a giant group of molecular clouds which is associated with the Sco OB2 association \citep{Loren}.  The dense region in the eastern part is known as the $\rho$ Oph main cloud, containing three B-type stars \citep{Yui}.  The $\rho$ Oph region is well studied by far-infrared forbidden lines \citep{Liseau,Yui}, H$_2$ \citep{Habart}, CO \citep{Nozawa,Loren}, and \ci~emission \citep{Kamegai}.  \citet{Liseau} made mapping observations of a wide area of the $\rho$ Oph main cloud with the LWS on board the {\it ISO} and derived physical properties of the PDR.  They suggested that the \oi~63\um~to 146\um~ratio is too small ($<5$) to be accounted for by PDR models. \citet{Habart} made high spatial-resolution observations of H$_2$ pure rotational and ro-vibrational transitions in the northern-western part with the SWS.  They derived that the ortho-to-para (OTP) ratio is close to unity and suggested that the H$_2$ formation rate must be high in the warm gas.  Here we report one-dimensional raster-scan observations with the SWS and LWS, which give us the spatial distribution of \siii~35\um~emission and the correlation with other line emissions from the PDR, and provide a constraint on the Si abundance in the PDR gas.

In addition, we observed a region around $\sigma$ Sco, a B1 type star located south-west of the $\rho$ Oph region in order to examine PDR properties and the Si abundance in the same association as $\rho$ Oph.  \citet{Baart} suggested the presence of a spherical \ion{H}{2} region around $\sigma$ Sco from the radio continuum emission.  \citet{Yui} detected extended \cii~emission with a small peak near $\sigma$ Sco.  \citet{deGeus} did not detect the CO line emission around $\sigma$ Sco, which indicates that the interstellar gas around $\sigma$ Sco is almost completely photodissociated.

\section{Observations and data reduction}

\begin{figure}
\epsscale{1.0}
\plotone{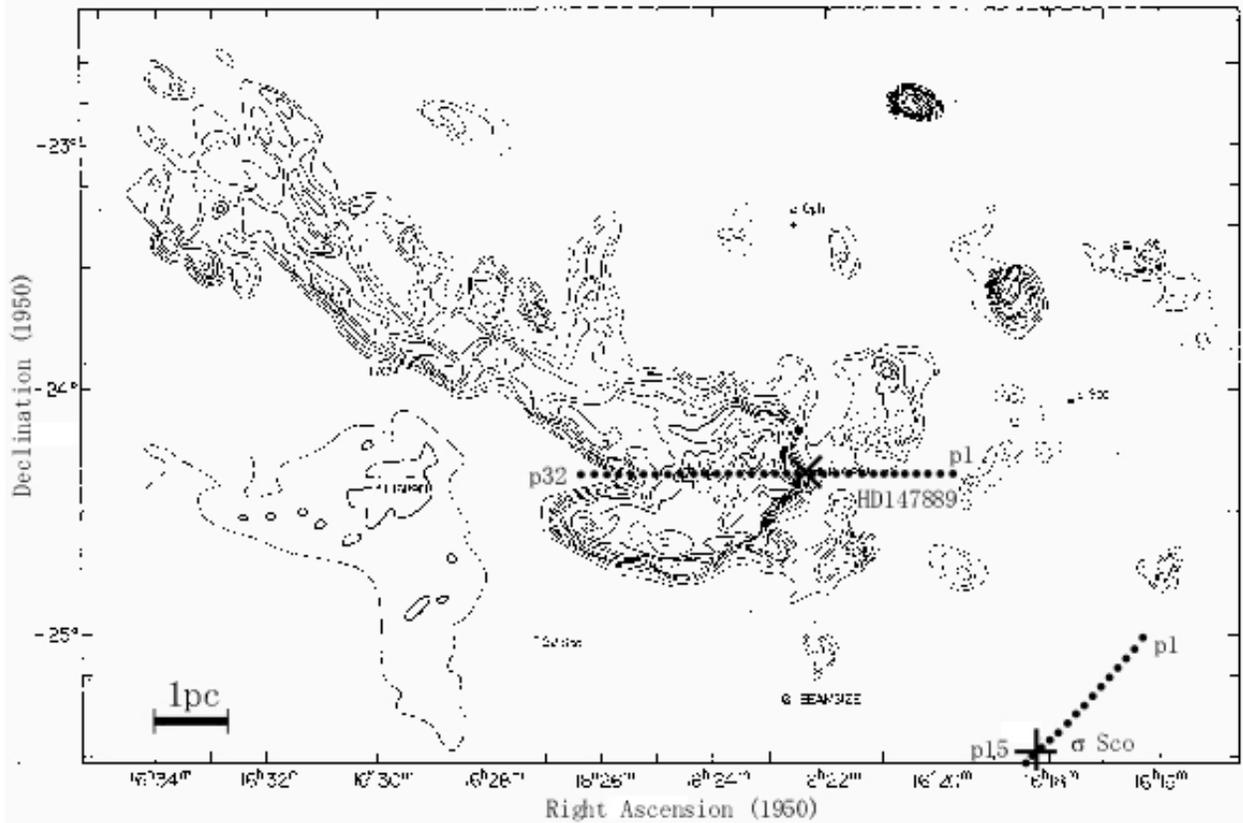}
\caption{Observed positions of the $\rho$ Oph and $\sigma$ Sco region.  The $^{13}$CO ($J=1$--0) peak temperature is shown in contours \citep{Loren}.  The positions of HD~147889 and $\sigma$ Sco star are shown by crosses.  The black circles show the observed positions and the size of the circles indicates the beam size of the LWS.\label{obsposition}}
\end{figure}

The observations were carried out at 32 positions in the $\rho$ Oph cloud (Fig.~\ref{obsposition} and Tables~\ref{rOphres_forbidden} and \ref{rOphres_H2}), each of which was separated by $3\arcmin$, or 0.12~pc at the distance of the $\rho$ Oph cloud of 136 pc \citep{Perryman}.  The position number (p\#) is designated to increase from west to east (toward the molecular cloud; see Fig.~\ref{obsposition}).  The exciting source HD~147889 is located between p13 and p14.  A peak of $^{13}$CO is located near p19.  For the $\sigma$ Sco region, one-dimensional raster-scan observations were carried out through $\sigma$ Sco toward the north-west direction, where the \cii~158\um~and the IRAS 100\um~continuum emissions show an extended structure \citep{Yui}.  The observations were carried out at 15 positions (Table~\ref{sScores}), each of which was separated by $3\arcmin$.  The position number (p\#) is designated from north-west to south-east.  The exciting source $\sigma$ Sco is located between p13 and p14 (see Fig.~\ref{obsposition}).  

For both regions, the line profile mode AOT SWS02 was used in the SWS observations to observe \siii~35\,$\mu$m and the H$_2$ pure rotational transition S(3) at 9.66\,$\mu$m in the cycle 1 observation.  The aperture size was $14\arcsec\times 20\arcsec$ and $20\arcsec\times 33\arcsec$, the spectral resolution $\lambda/\Delta\lambda$ was $1500$ and $2000$, and the flux calibration accuracy for point-like sources was 7\% and 22\% \citep{Leech} for H$_2$ S(3) and \siii~35\um~emissions, respectively.  Further observations were made from p13 to p20 for other H$_2$ pure rotational and ro-vibrational transitions in the $\rho$ Oph region in the cycle 2 (Table~\ref{rOphres_H2}).  The aperture size for each transition line observation is not the same (Table~\ref{rOphres_H2}).  The Off-Line Processing (OLP) version 10.1 data obtained from the ISO Archival Data Center were used for the present study.  At each raster position one up-and-down grating scan was carried out.  The spectra were further processed by using the ISO Spectral Analysis Package (ISAP\footnote{The ISO Spectral Analysis Package (ISAP) is a joint development by the LWS and SWS Instrument Teams and Data Centers.  Contributing institutes are CESR, IAS, IPAC, MPE, RAL and SRON.}) in the same manner as for S171 (OOSD03).  The conversion factors for extended sources are taken from \citet{Leech}, which are estimated to be accurate within 10\%.  The line intensities were derived by Gaussian fits. 

The same positions were observed with the LWS full grating scan mode AOT LWS01 to cover the wavelength range from 43 to 197\,$\mu$m with $\lambda/\Delta\lambda\sim 100$--$300$ \citep{Gry}.  Four grating scans were carried out for each raster position with the spectral sampling of half the resolution element.  The total integration time at a grating position was 2 sec.  The LWS beam size was $66\arcsec$--$86\arcsec$, depending on the wavelength \citep{Gry}.  The OLP 10.1 data were used in the present study.  We used the ISAP for further data processing and derived the line intensities by Gaussian fits.  The absolute flux calibration uncertainty is reported to be 10--20\% for point sources and 50\% for extended sources \citep{Gry}.

\section{Results and Discussions}

In the $\rho$ Oph region, we detected 2 lines, H$_2$ S(3) 9.66\um~and \siii~35\um~in the SWS spectra and 3 forbidden lines, \oi~63\um, \oi~146\um, and \cii~158\um~in the LWS spectra for the cycle 1 observations.  In the cycle 2 we detected three pure rotational lines of H$_2$, S(2) 12.3\um, S(1) 17.0\um, and S(0) 28.2\um.  The results are summarized in Tables~\ref{rOphres_forbidden} and \ref{rOphres_H2}.  The detected ionic lines arise only from elements with ionization potential less than that of hydrogen (C and Si), i.e., no higher ionization lines, such as \nii~122\um~nor \oiii~88\um, have been detected, being consistent with the absence of a radio continuum emission peak around HD~147889 \citep{Baart}.  The \oi~146\um~line was observed by two adjacent detectors, LW3 and LW4.  The line intensities derived from different detectors are in agreement with each other within the estimated uncertainties.  We use the weighted mean of the intensities from the two detectors in the following analysis.  In the case that the line was detected only by one detector with the other being an upper limit, we adopted the intensity of the detected channel.  

The results of the $\sigma$ Sco region are summarized in Table~\ref{sScores}.  \siii~35\um~and H$_2$ 9.66\um~have not been detected at any positions in the SWS spectra.  In the LWS spectra we detected 3 forbidden lines, \oi~63\um, \nii~122\um, and \cii~158\um.  The \nii~122\um~emission traces the low density ionized gas and the detection of this line is consistent with the presence of the radio continuum emission peak \citep{Baart}.  On the other hand, the lack of emission lines from highly ionized ions such as \oiii~and \niii~is consistent with the late spectral type (B1) of the exciting star. 

The errors in the line intensities include both the fitting errors and the statistical errors of the baseline.  Uncertainties in the absolute flux are not included in the errors in Tables~\ref{rOphres_forbidden}--\ref{sScores}.  For the line emission less than 3 times the uncertainty ($<3\sigma$) we regard it as non-detection and give $3\sigma$ as the upper limit.  For each region, the lines not listed in Tables~\ref{rOphres_forbidden}--\ref{sScores} are not detected at all the observed positions and only upper limits are derived as: \oiii~52\um\,$<2.7\times 10^{-7}$, \niii~57\um\,$<2.0\times 10^{-7}$, \oiii~88\um\,$<6.9\times 10^{-8}$ and \nii~122\um\,$<3.6\times 10^{-8}$\,W\,m$^{-2}$\,sr$^{-1}$ for the $\rho$ Oph region, H$_2$ 9.66\um\,$<3.6\times 10^{-8}$, \siii~35\um\,$<6.4\times 10^{-8}$, \oiii~52\um\,$<5.3\times 10^{-8}$, \niii~57\um\,$<4.9\times 10^{-8}$, \oiii~88\um\,$<3.2\times 10^{-8}$, and \oi~146\um\,$<3.7\times 10^{-9}$\,W\,m$^{-2}$\,sr$^{-1}$ for the $\sigma$ Sco region.

\begin{figure}
\epsscale{1.0}
\plotone{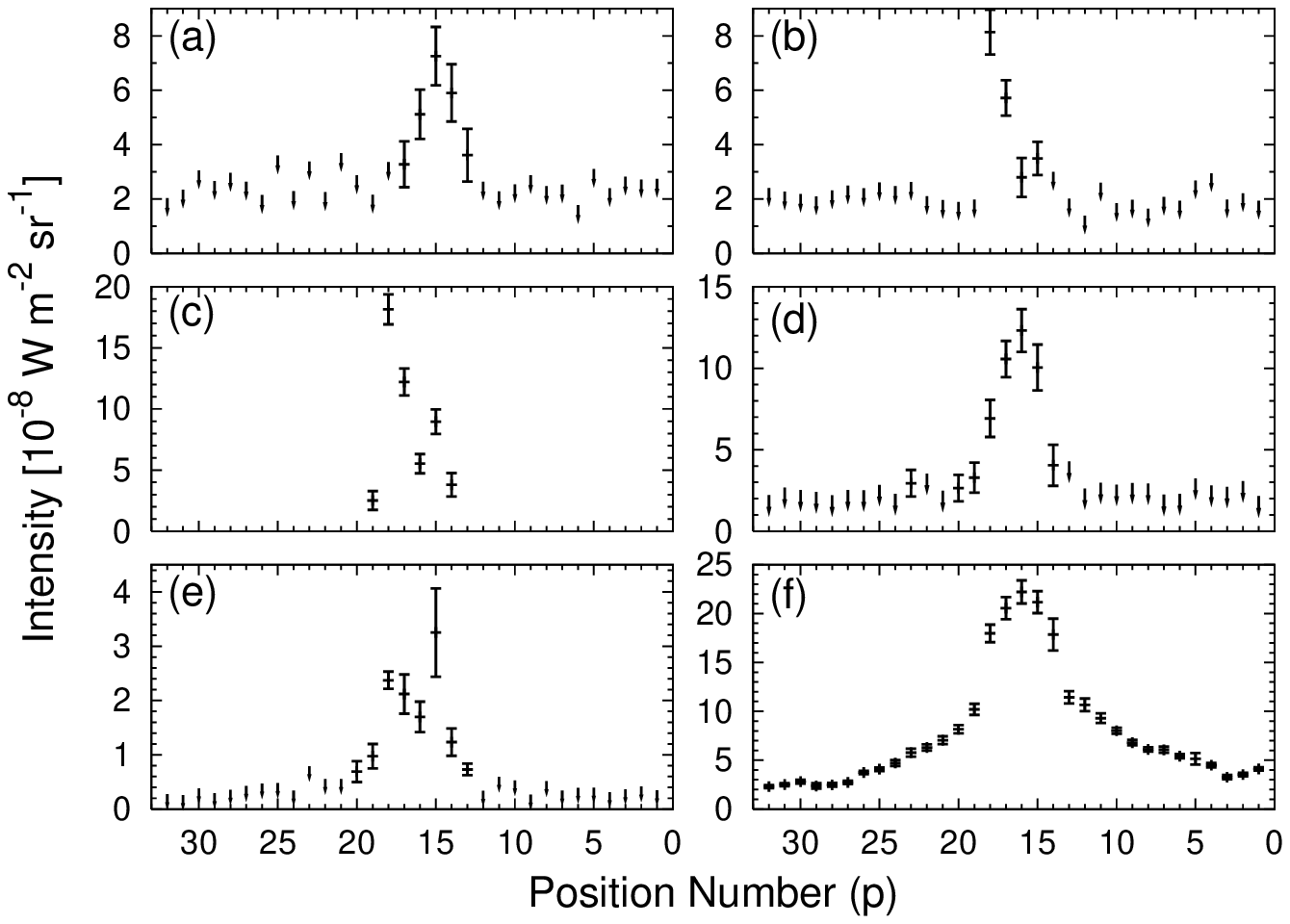}
\caption{Observed line intensities in the $\rho$ Oph region.  HD~147889 is located between p13 and p14.  The arrows indicate 3$\sigma$ upper limits.  {\bf a)} \siii~35\um;  {\bf b)} H$_2$ 9.66\um;  {\bf c)} H$_2$ 17.0\um;  {\bf d)} \oi~63\um;  {\bf e)} \oi~146\um;  and {\bf f)} \cii~158\um. \label{rOphlines}}
\end{figure}

\begin{figure}
\epsscale{1.0}
\plotone{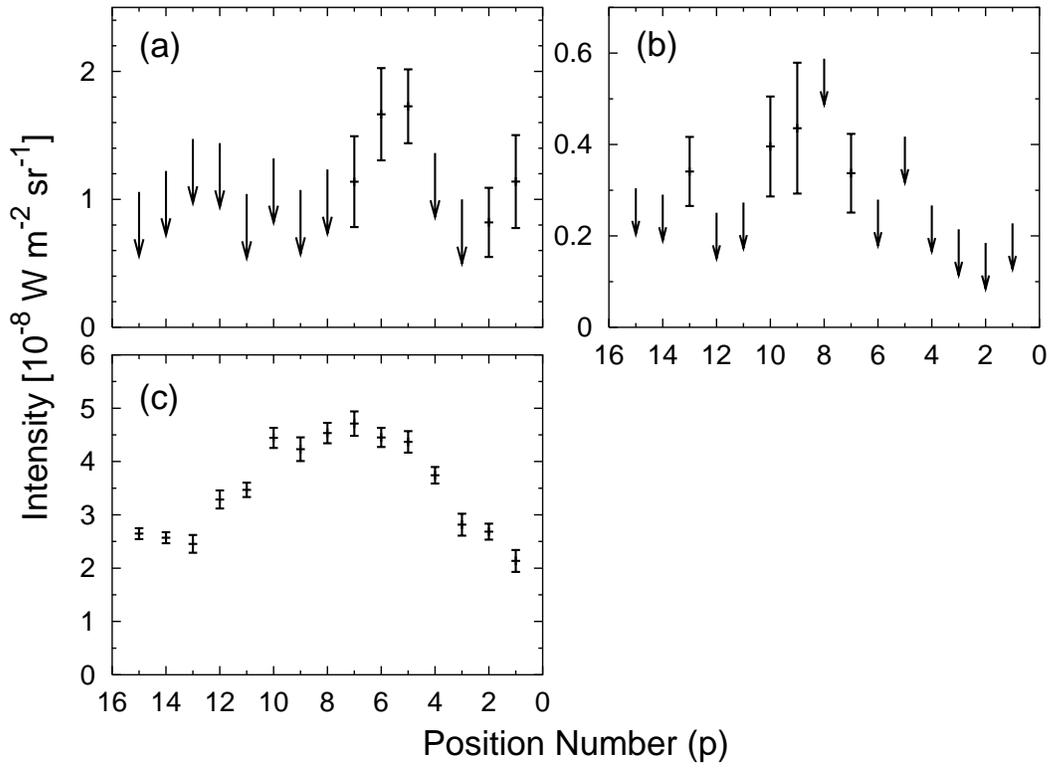}
\caption{Observed line intensities in the $\sigma$ Sco region.  $\sigma$ Sco is located between p13 and p14.  The arrows indicate 3$\sigma$ upper limits.  {\bf a)} \oi~63\um;  {\bf b)} \nii~122\um; and {\bf c)} \cii~158\um. \label{sScolines}}
\end{figure}

Figures~\ref{rOphlines} and \ref{sScolines} show the spatial distributions of the line intensities for the $\rho$ Oph and the $\sigma$ Sco regions.  For the $\rho$ Oph region, the line emissions other than \cii~158\um~are detected only at p13--p20 and p23, near the surface regions of the molecular clouds.  The \cii~158\um~emission is detected at all the observed positions and shows a peak at p15.  This is consistent with the result of \citet{Yui}, who reported an extended distribution of the \cii~emission.

\subsection{PDR properties}

We estimate the properties of the PDR from the observed lines and continuum emission using the PDR model by \citet{Kaufman}.  The major model parameters are the gas density, $n$, and the UV radiation field strength, $G_0$, in units of the solar neighborhood value \citep[$1.6\times 10^{-6}\,\mbox{W{\,}m}^{-2}$]{Habing}.  The model gives integrated line intensities seen in the face-on view.  In the following, we discuss only the results of the positions where all the \oi~63\um, 146\um, and \cii~158\um~emissions are detected (p14--p20).  Although any appreciable amount of the ionized gas has not been detected in the $\rho$ Oph region, part of \cii~158\um~may come from the undetected ionized gas.  From 2.3GHz radio continuum observations \citep[Emission measure $=176$~pc cm$^{-6}$;][]{Baart} we estimate the contribution of the \cii~158\um~emission from the ionized gas to be less than 6\% assuming the low-density limit condition and the interstellar abundance \citep{SavageSembach}.  This fraction is smaller than the uncertainty in the observed intensity.  Possible contribution from the ionized gas thus will not affect the following discussion on the PDR properties within the observational uncertainties.  Comparison between the observation and the PDR model can be made less ambiguously for the $\rho$ Oph region than for \ion{H}{2} regions, such as S171 (OOSD03).

\begin{figure}
\epsscale{1.0}
\plotone{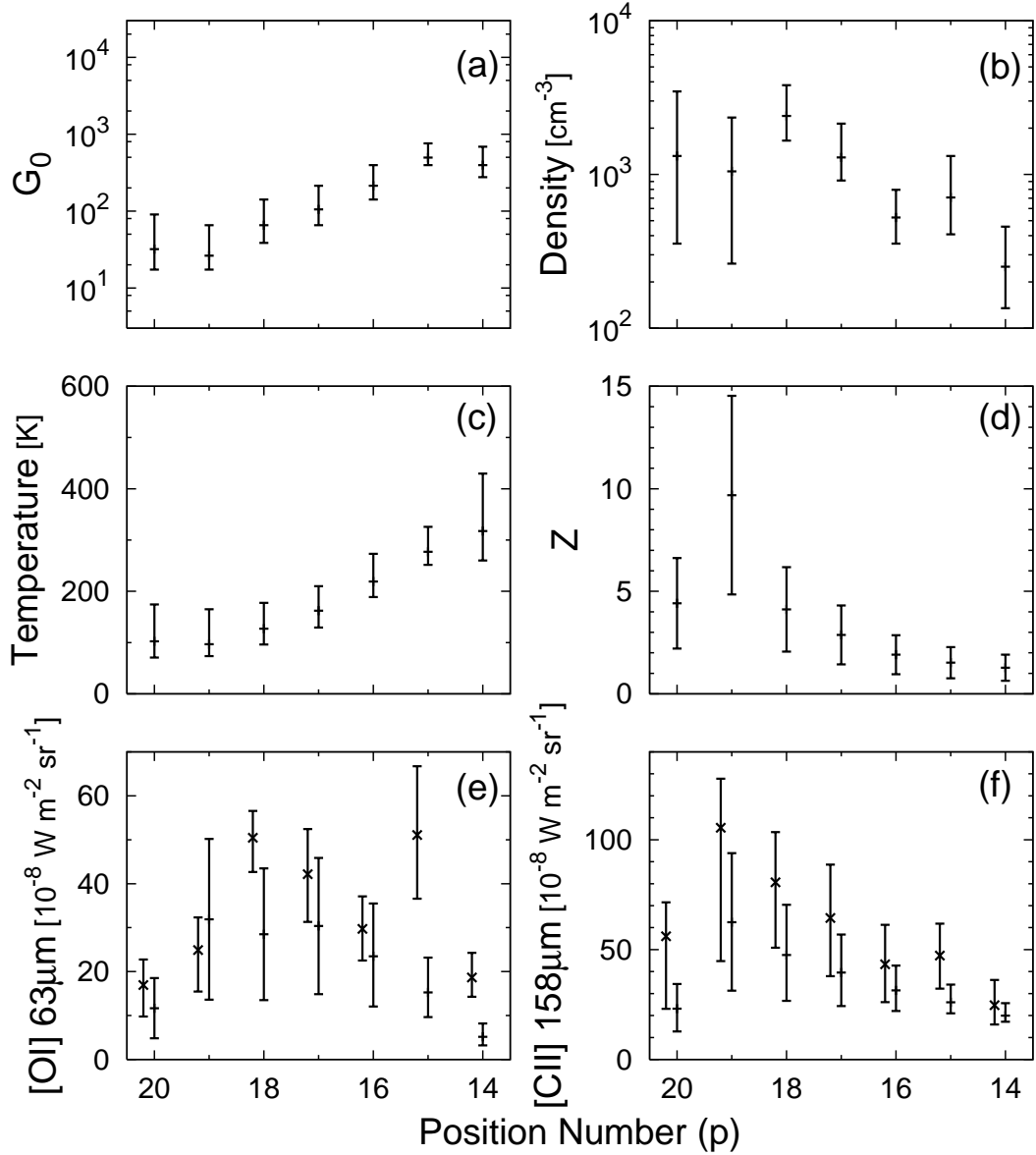}
\caption{PDR properties for the $\rho$ Oph region. {\bf a)} Radiation field strength $G_0$ estimated from the continuum of the LWS spectrum (see text), {\bf b)} gas density, {\bf c)} surface temperature from the PDR model, {\bf d)} overlapping factor $Z$ (see text), {\bf e)} model prediction of the \oi~63\um~intensity derived from the PDR model (crosses shifted slightly to left for clarity) and the observed intensity of \oi~63\um~after correcting the overlapping effect using $Z$ (pluses) (see text), and {\bf f)} same as e) but for \cii~158\um.  \label{rOphPDRres}}
\end{figure}

The ratio of the observed \oi~63\um/\oi~146\um~is too small (3--7) compared to PDR models of any parameter range, and no PDR models can account for the observed line intensities of \oi~63\um, 146\um, and \cii~158\um~consistently.  The small ratio of \oi~63\um/\oi~146\um~in the $\rho$ Oph region has been reported by \citet{Liseau} and is attributed to the optically thick emission of \oi~63\um~\citep{Caux}.  To derive the physical properties of the PDR and explain quantitatively the discrepancy between the observed line intensities and the prediction from the PDR model by \citet{Kaufman}, we adopt a simple model that was applied for S171 (OOSD03).  \oi~63\um~is optically thick, \cii~158\um~is marginally thick ($\tau\sim 1$), and \oi~146\um~and the FIR continuum emission are optically thin in a PDR \citep{Kaufman}.  OOSD03 suggest that overlapping PDR clouds along the line of sight attenuate the former two line emissions, while the intensities of the latter two will be a simple summation of those from each PDR cloud.  Therefore, we use the \oi~146\um~to the total far-infrared intensity ($FIR$) ratio and $G_0$ as input parameters to estimate the density and the surface temperature using the PDR model (Figs.~\ref{rOphPDRres}a--c).   We fit the continuum emission of the LWS spectra with a graybody radiation of the $\lambda^{-1}$ emissivity and estimate $G_0$ by assuming $\langle \tau_{\mathrm{abs}} \rangle /\tau_{100}=700$ and $f=0.5$, where $\langle \tau_{\mathrm{abs}} \rangle$ is the weighted mean of the absorption optical depth over the mean intensity from the heating source, $\tau_{100}$ is the optical depth at 100\um, and $f$ is the ratio of the energy between 6~eV and 13.6~eV to the total luminosity for the spectrum of HD~147889 of a blackbody of $T_{\mathrm{eff}}=20300$~K (see OOSD03 for details).  Models with these parameters predict the \oi~63\um~and \cii~158\um~emissions stronger than observed.  It can be explained by taking into account of the overlapping factor $Z$ defined as
\begin{equation}
Z=\frac{FIR(\mathrm{obs})}{FIR(\mathrm{model})},
\end{equation}
where $FIR(\mathrm{model})=1.6\times 10^{-6}\,G_0/4\pi f\ \mathrm{W\,m}^{-2}\mathrm{\,sr}^{-1}$ with the assumption that all the incident radiation is absorbed and converted into the FIR emission.  $Z$ is equal to the beam filling factor for $Z<1$, whereas it indicates the degree of overlapping of clouds for $Z>1$ since the FIR emission is optically thin.  Figure~\ref{rOphPDRres}d shows the estimated $Z$ with the errors including the absolute flux uncertainty of the LWS.  At the positions of p14--p20, $Z$ is larger than unity, suggesting that the effect of overlapping clouds is significant.  We assume that the size of a cloud is smaller than the beam ($\lesssim 0.06$~pc), and the column density for \oi~63\um~in a cloud is sufficiently large ($N_{\mathrm{H}+\mathrm{H}_2} \gtrsim 5\times 10^{21}$~cm$^{-2}$).  To examine the effect of the overlapping on the observed intensity we simply assume that $N$ identical clouds overlap on the line of sight and attenuate the line emissions.  Then the observed \oi~63\um~and \cii~158\um~should be attenuated by a factor of
\begin{equation}
\frac{\sum_{j=0}^{N-1}e^{-j\tau}}{N}=\frac{1-e^{-N\tau}}{N(1-e^{-\tau})}\,,
\end{equation}
where the number of clouds $N$ can be substituted by $Z$ (OOSD03).  We assume $\tau=\infty$ for \oi~63\um~and $\tau=1$ for \cii~158\um.  The corrected \oi~63\um~and \cii~158\um~intensities together with the observed \oi~146\um~and the FIR emissions can be accounted for by the PDR model of $G_0$ and $n$ of Figs~\ref{rOphPDRres}a and b within $1\sigma$ except for p14 and p15, where the model prediction of the \oi~63\um~is still slightly larger than the observed intensities (Figs.~\ref{rOphPDRres}e and f).  In Figs.~\ref{rOphPDRres}e and f, the errors in the model prediction include the uncertainty in the derived parameters ($G_0$ and $n$), whereas those in the observed intensities include the uncertainties in the original intensity (Table~\ref{rOphres_forbidden}) and those in $Z$ (Fig.~\ref{rOphPDRres}d).  The derived distributions of $G_0$, $n$ and the surface temperature are compatible with the CO observations and the location of the exciting source.  In the present estimate, we assume that the cloud density is the same among PDRs along the line of sight to simplify the calculation and the derived density represents a mean value of the true distribution.  The agreement of the model prediction with the corrected line intensities at 5 positions supports interpretation by overlapping PDR clouds for the line intensities in a semi-quantitative way.

\begin{figure}
\epsscale{1.0}
\plotone{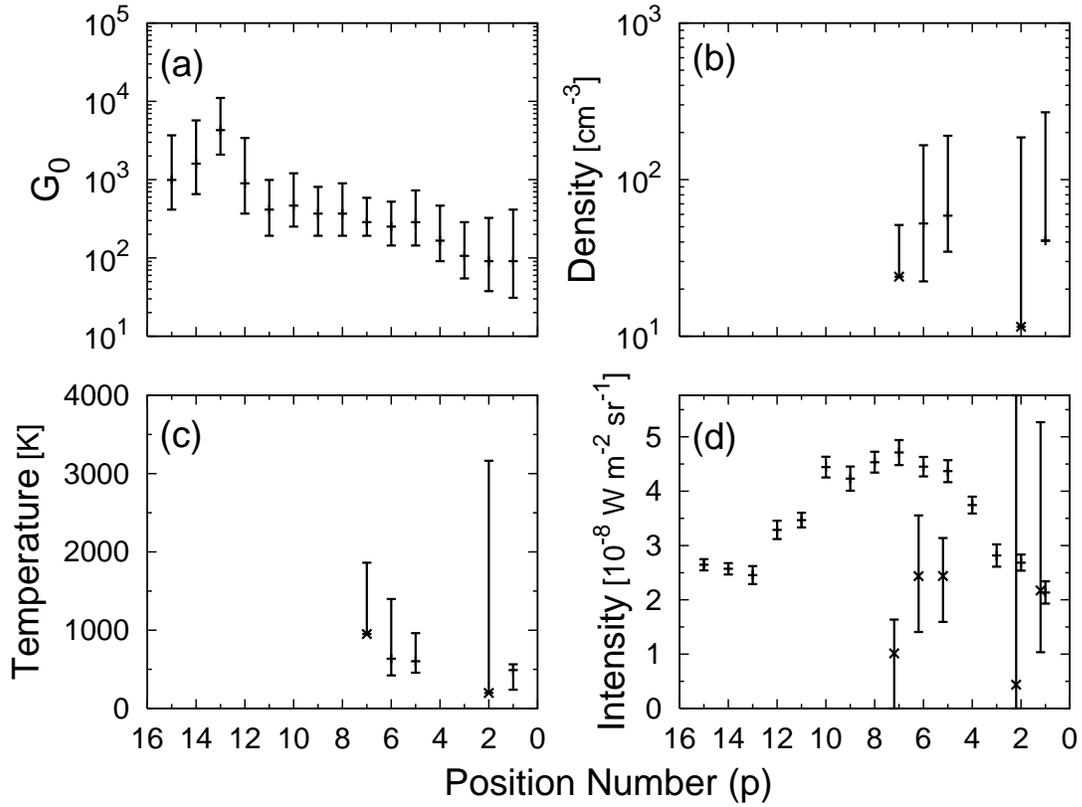}
\caption{PDR properties for the $\sigma$ Sco region. {\bf a)} Radiation field strength $G_0$ estimated from the continuum of the LWS spectrum (see text), {\bf b)} gas density, {\bf c)} surface temperature, and {\bf d)} the model prediction of the \cii~158\um~intensity derived from the PDR model (crosses shifted slightly to left for clarity). Pluses indicate the observed \cii~158\um~intensitiy. \label{sScoPDRres}}
\end{figure}

For all the 15 observed positions in the $\sigma$ Sco region we find $Z<1$, which indicates that the \oi~63\um~emission is unlikely to be attenuated by overlapping PDR clouds.  Therefore, we use the \oi~63\um~emission instead of the \oi~146\um~emission to derive the PDR properties (Figs.~\ref{sScoPDRres}a--c).  $\sigma$ Sco is associated with a \ion{H}{2} region and the \cii~158\um~emission originates both from the ionized and the PDR gas.  The \cii~158\um~intensity predicted by the PDR model is smaller than those observed (Fig.~\ref{sScoPDRres}d).  The excess intensity is attributed to the emission from the ionized gas.  The contribution from the ionized gas is 30--80\% at three positions closest to the excitation star (p5, p6, and p7).  \citet{Yui} suggested that the \cii~158\um~emission comes from the neutral gas adjacent to the \ion{H}{2} region based on the poor correlation between the radio continuum and the \cii~158\um~emission.  The present results indicate that a non-negligible fraction of the \cii~158\um~emission comes from the ionized gas in the $\sigma$ Sco region.

\subsection{Silicon abundance in the PDR}

\begin{figure}
\epsscale{1.0}
\plotone{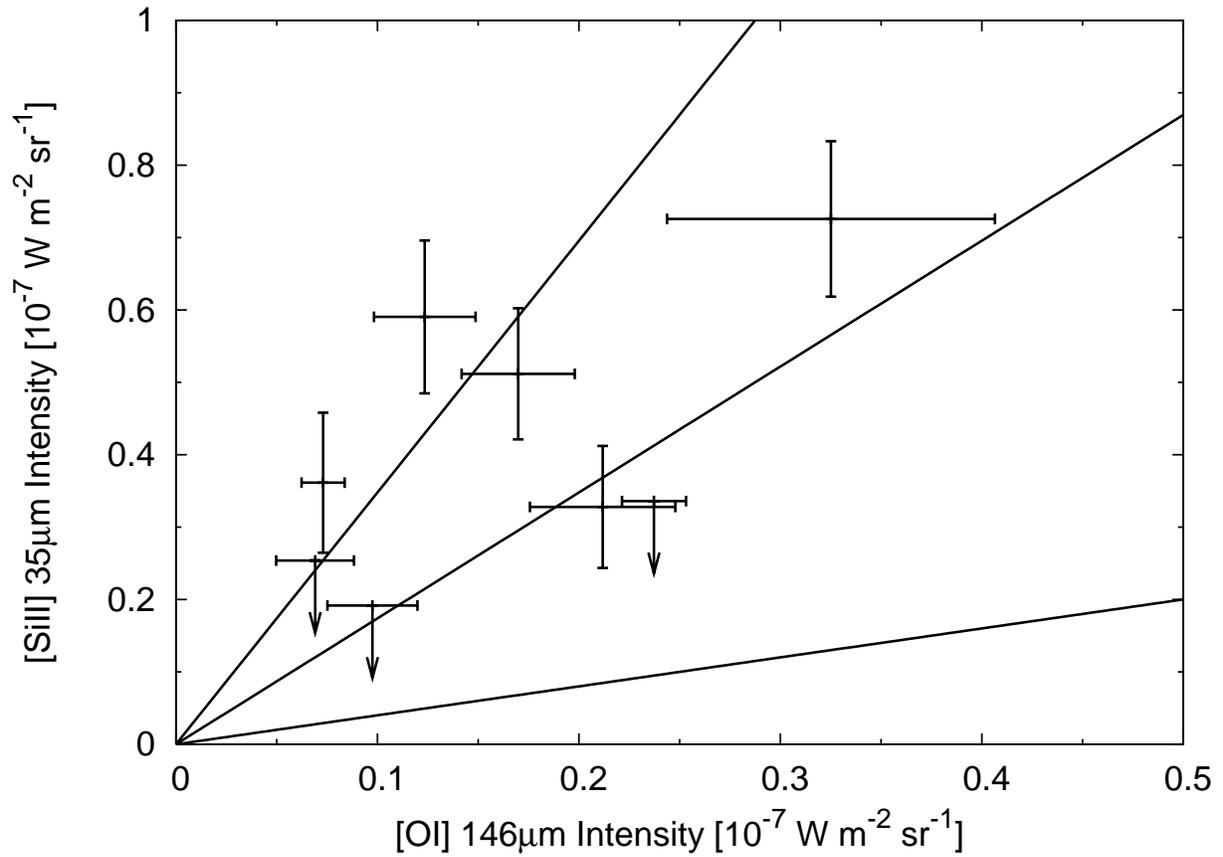}
\caption{The observed \oi~146\um~against \siii~35\um~line intensities at positions where at least one of them is detected.  The solid lines show the Si abundance of 2.3\% (PDR model), 10\%, and 20\% of solar from bottom to top. \label{OI146SiII}}
\end{figure}

In the $\rho$ Oph region, we have detected strong \siii~35\um~emission.  With the gas density $\sim 10^3$~cm$^{-3}$ and $G_0\sim 10^3$, the PDR model \citep{Hollenbach} predicts the \siii~35\um~to the \oi~146\um~intensity ratio to be $\sim 0.4$, while the observation shows the ratio of $\sim 2$--$4$ (Fig.~\ref{OI146SiII}).  The abundance of Si is assumed to be 2.3\% of the solar abundance in the model.  If we attribute the difference to the abundance or the depletion degree of Si, it suggests that 10--20\% of the solar abundance of Si must reside in the gas phase in the PDR of $\rho$ Oph.  The optical depth of \siii~35\um~in a typical PDR is about 0.1 and thus the overlapping effect is negligible.  The \siii~35\um~and \oi~146\um~transitions have similar excitation energies (413~K and 327~K, respectively) and both transitions as well as \oi~63\um~have critical densities higher than $10^5$~cm$^{-3}$ with the collision partner of atomic hydrogen.  The calculation for isothermal cloud with a uniform density with the optically thin condition shows that the \siii~35\um~to the \oi~146\um~ratio is not dependent on the gas density for $n\lesssim 10^4$~cm$^{-3}$ and not sensitive to the gas temperature for $T\gtrsim 300$~K.  In fact PDR model calculations suggest that the ratio changes only by 10\% for $n<10^4$~cm$^{-3}$ and $10^2 \lesssim G_0 \lesssim 10^3$ \citep{Wolfire}.  
\citet{Meixner} showed that clumpy PDRs can increase the \siii~35\um~to \oi~146\um~intensity ratio by a factor of 1.5.  Clumpiness alone can not fully account for the observed large ratio.  \citet{Liseau} showed that the density of the PDR in the $\rho$ Oph main cloud is (1--3)$\times 10^4$~cm$^{-3}$, which is by one order of magnitude larger than the present result.  The discrepancy comes from the difference of the employed PDR models: the model by \citet{Kaufman} takes account of the heating by PAHs that produces strong \oi~emissions at low densities.  The \siii~35\um~to the \oi~146\um~intensity ratio is not sensitive to the density as mentioned above, and thus the large ratio must be attributed to the large Si abundance irrespective of the assumed density.  As the \cii~158\um~emission, the \siii~35\um~emission can originate from the ionized gas.  We estimate from the radio continuum observation \citep{Baart} that only 30\% of the observed \siii~35\um~emission can be attributed to the ionized gas even if the gaseous Si abundance is solar.  Therefore most \siii~35\um~emission should come from the PDR and the possible contribution from the ionized gas will not affect the above discussion on the large gas phase abundance of Si in the PDR.

\begin{figure}
\epsscale{1.0}
\plotone{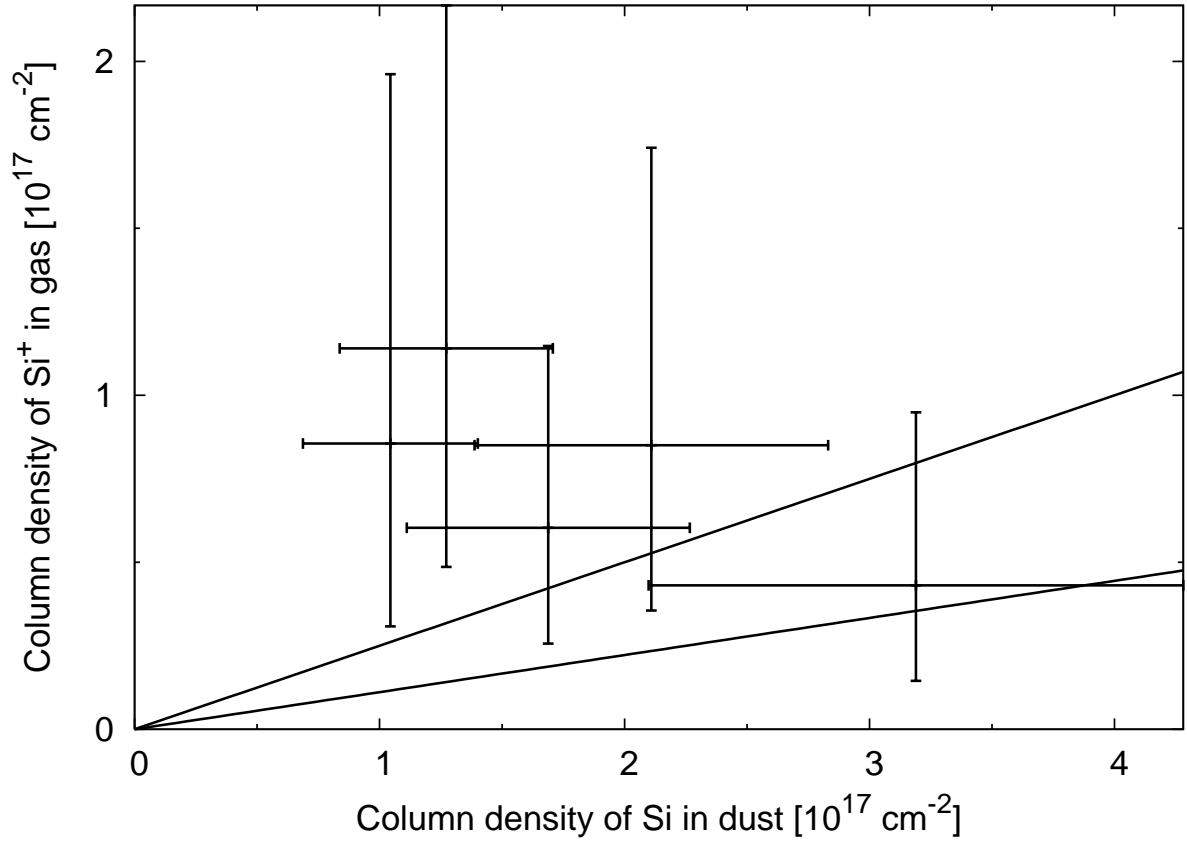}
\caption{Column density of Si in dust against that in gas.  Only the data at p13-p17 are plotted, where the \siii~35\um~emission is detected.  Solid lines show the relative abundance of Si in gas : dust $=$ 1:9 (10\% is in gas phase) and 1:4 (20\% is in the gas phase) from bottom to top. \label{Sigasdust}}
\end{figure}

The column density of Si included in dust grains can be estimated from the dust model and $\tau_{100}$, which is derived from the graybody fit of the continuum spectra.  We use the dust model by \citet{DraineLee} with the assumptions of the density of dust grains of 3.3\,g\,cm$^{-3}$ and the mean molecular weight of 100--200 for olivine and pyroxene.  On the other hand, we can estimate the column density of Si$^+$ from the \siii~35\um~intensity, the gas density, and the gas temperature.  We use the density and the surface temperature derived from the PDR model in \S 3.1 in the following discussion.  For a given intensity, the column density decreases with the temperature and thus a highest likely temperature gives a lower limit of the column density.  Figure~\ref{Sigasdust} plots the column density of Si in gas against that in dust.  It indicates that the relative fraction of Si in the gas and dust phases is in agreement with the result from the \siii~35\um/\oi~146\um~ratio.  Note that this estimate is independent of the assumed solar abundance.  If the density is higher by one order of magnitude as suggested by \citet{Liseau}, we obtain a lower column density of Si in the gas phase by one order of magnitude.  Then \siii~35\um/\oi~146\um~is estimated to be lower than observed.  The gas phase abundance of Si estimated from \siii~35\um~and $\tau_{100}$ supports the present estimate of the gas density ($\sim 10^3$~cm$^{-3}$), though we cannot exclude the higher density because of large uncertainties.

The temperature of H$_2$ emitting gas can be estimated from the ratio of the H$_2$ line emissions.  Assuming the Boltzmann distribution at low-J transitions and the homogeneous distribution of the emissions in the aperture we derive the temperature of ortho H$_2$ to be 298$\pm$12~K, 315$\pm$20~K, 310$\pm$9~K, and 307$\pm$8~K at p15--p18 in the $\rho$ Oph region, respectively, from the line intensities of 9.66\um~and~17.0\um.  Although the aperture size for each line is not the same (Table~\ref{rOphres_H2}), similar temperatures derived at 4 positions suggest that the differences in the aperture size do not significantly affect the estimates.  At p18, we also detect two lines from para H$_2$ at 12.3\um~and 28.2\um, and derive the temperature as 260$\pm$21~K, which gives a similar temperature as that from ortho H$_2$ lines.  The temperature of the H$_2$ emitting gas is in a range similar to the surface temperature.  If we use the temperature of H$_2$ emission to estimate the column density of gaseous Si$^+$, the conclusion remains unchanged.

According to theoretical investigations, dust grains in the interstellar space are destroyed in a time scale of ($2$--$4$)$\times10^8$ yr by supernova shock waves \citep{Jones94,Jones96}.  However, the gas densities of the PDR are several hundreds to several thousands cm$^{-3}$, and a shock can not penetrate these regions easily.  Some fraction of Si may reside in volatile forms in dust grains which can be easily destroyed.  Several forms of Si in dust grains are proposed to explain the fact that an order of 10\% of Si atoms easily return to the gas phase: a volatile component with the binding energy of $E_b\simeq 1$--$2$~eV \citep{Tielens}, a mantle being photodesorbed by UV photons \citep{Walmsley}, and organometallic complexes \citep{Klotz,Marty}.  The present study suggests that about 10\% of Si atoms may reside in some of these forms in dust grains that are relatively easily ejected to the gas phase by UV photons.  The recycling as well as the composition of dust grains in interstellar space should be reexamined based on further systematic observations of variations in different environments and other dust composing elements.

In the $\sigma$ Sco region, the upper limit of the \siii~35\um~to \nii~122\um~intensity ratio indicates that the Si abundance in the ionized gas should be lower than 100\% of the solar abundance and no useful constraint on the abundance of Si in gas phase is obtained.

\section{Summary}

We observed the $\rho$ Oph main cloud and the $\sigma$ Sco region with SWS and LWS on board the {\it ISO}.  We have detected 8 lines, \siii~35\um, H$_2$ pure rotational transition lines S(0) to S(3), \oi~63\um, 146\um, and \cii~158\um, in the $\rho$ Oph region, and 3 fine-structure lines, \oi~63\um, \nii~122\um, and \cii~158\um, in the $\sigma$ Sco region.

In the $\rho$ Oph region, the observed \oi~146\um~emission is too strong, and the \oi~63\um~and the \cii~158\um~emissions are too weak compared to PDR models.  As has been proposed for S171 (OOSD03), we adopt a simple model of overlapping PDR clouds along the line of sight which attenuate the \oi~63\um~and \cii~158\um.  We show that this model accounts for the intensities of the \oi~63\um, 146\um, \cii~158\um, and the continuum emission at the same time.  Because of the absence of the \ion{H}{2} region in the $\rho$ Oph, the present analysis supports the validity of the overlapping model with less ambiguity.  The gas density varies over $250$--$2500$~cm$^{-3}$ and the surface temperature $T\sim 100$--$400$~K for the $\rho$ Oph region. In the $\sigma$ Sco region, the overlapping effect is negligible and we derive physical conditions from the \oi~63\um~emission.  Comparison of the model prediction indicates that 30--80\% of the \cii~158\um~emission comes from the ionized gas in the vicinity of $\sigma$ Sco.

The \siii~35\um~emission was observed quite strongly in the $\rho$ Oph region.  With the derived PDR properties we discuss the large Si abundance in the PDR gas in the $\rho$ Oph region.  The observed ratio of \siii~35\um~to the \oi~146\um~indicates the Si abundance in the PDR gas to be 10--20\% of solar, suggesting the dust destruction taking place in the PDR.  This result is supported by comparison of the \siii~35\um~intensity with the continuum emission for $n\sim 10^3$~cm$^{-3}$.  The present results strongly suggest that some fraction of Si atoms must be included in volatile forms in dust grains which are relatively easily destroyed.  The present observation indicates the importance of the \siii~35\um~line in future observations to investigate the dust destruction and the composition of dust grains.

\acknowledgments
The authors thank all the members of the Japanese ISO group, particularly H. Okuda,  K. Kawara, and Y. Satoh for their help during the observations and continuous encouragement.  They also thank F. Boulanger and K. Mochizuki for useful discussions.  This work was supported in part by Grant-in-Aids for Scientific Research from the Japan Society for the Promotion of Science (JSPS).

\clearpage

\begin{deluxetable}{ccrrccccc}
\tablecaption{Summary of the forbidden-line intensities in the $\rho$ Oph region \label{rOphres_forbidden}}
\tablehead{
\colhead{No.}  & \multicolumn{2}{c}{R.A.\tablenotemark{a}} & \colhead{DEC.\tablenotemark{b}} & \colhead{$d$\tablenotemark{c}} &
\multicolumn{4}{c}{line intensities ($10^{-8}\,\mbox{W{\,}m}^{-2}\,\mbox{sr}^{-1}$)}\\
\colhead{}& \colhead{m} & \colhead{s} & \colhead{s} & \colhead{pc} & \colhead{\siii~35\um} & \colhead{\oi~63\um} & \colhead{\oi~146\um} & \colhead{\cii~158\um}
}
\startdata
 1 & 22 & 37.15 & 39.83 & 1.50 & $<$ 2.73         &  $<$ 2.15         & $<$ 0.34         &  4.12 $\pm$ 0.17 \\ 
 2 & 22 & 50.33 & 41.17 & 1.39 & $<$ 2.72         &  $<$ 3.07         & $<$ 0.42         &  3.54 $\pm$ 0.17 \\ 
 3 & 23 &  3.52 & 42.45 & 1.27 & $<$ 2.82         &  $<$ 2.73         & $<$ 0.36         &  3.26 $\pm$ 0.21 \\ 
 4 & 23 & 16.70 & 43.66 & 1.15 & $<$ 2.40         &  $<$ 2.81         & $<$ 0.31         &  4.50 $\pm$ 0.22 \\ 
 5 & 23 & 29.88 & 44.80 & 1.03 & $<$ 3.10         &  $<$ 3.24         & $<$ 0.39         &  5.15 $\pm$ 0.59 \\ 
 6 & 23 & 43.07 & 45.86 & 0.91 & $<$ 1.77         &  $<$ 2.29         & $<$ 0.39         &  5.42 $\pm$ 0.20 \\ 
 7 & 23 & 56.24 & 46.85 & 0.79 & $<$ 2.53         &  $<$ 2.24         & $<$ 0.34         &  6.07 $\pm$ 0.32 \\ 
 8 & 24 &  9.43 & 47.77 & 0.67 & $<$ 2.47         &  $<$ 2.93         & $<$ 0.51         &  6.11 $\pm$ 0.22 \\ 
 9 & 24 & 22.60 & 48.62 & 0.56 & $<$ 2.87         &  $<$ 2.96         & $<$ 0.27         &  6.80 $\pm$ 0.27 \\ 
10 & 24 & 35.79 & 49.41 & 0.44 & $<$ 2.54         &  $<$ 2.86         & $<$ 0.53         &  8.02 $\pm$ 0.31 \\ 
11 & 24 & 48.97 & 50.11 & 0.32 & $<$ 2.28         &  $<$ 2.97         & $<$ 0.59         &  9.30 $\pm$ 0.48 \\ 
12 & 25 &  2.15 & 50.75 & 0.20 & $<$ 2.63         &  $<$ 2.63         & $<$ 0.34         & 10.66 $\pm$ 0.64 \\ 
13 & 25 & 15.34 & 51.31 & 0.08 &  3.61 $\pm$ 0.97 &  $<$ 4.29         &  0.73 $\pm$ 0.11 & 11.43 $\pm$ 0.63 \\ 
14 & 25 & 28.52 & 51.80 & 0.04 &  5.90 $\pm$ 1.06 &   4.05 $\pm$ 1.25 &  1.23 $\pm$ 0.25 & 17.85 $\pm$ 1.63 \\ 
15 & 25 & 41.70 & 52.22 & 0.16 &  7.26 $\pm$ 1.07 &  10.05 $\pm$ 1.41 &  3.25 $\pm$ 0.81 & 21.17 $\pm$ 1.12 \\ 
16 & 25 & 54.88 & 52.57 & 0.28 &  5.12 $\pm$ 0.91 &  12.32 $\pm$ 1.31 &  1.78 $\pm$ 0.32 & 22.21 $\pm$ 1.19 \\ 
17 & 26 &  8.06 & 52.85 & 0.39 &  3.28 $\pm$ 0.84 &  10.57 $\pm$ 1.11 &  2.12 $\pm$ 0.36 & 20.55 $\pm$ 1.13 \\ 
18 & 26 & 21.25 & 53.06 & 0.51 & $<$ 3.36         &   6.92 $\pm$ 1.14 &  2.37 $\pm$ 0.16 & 17.97 $\pm$ 0.90 \\ 
19 & 26 & 34.43 & 53.20 & 0.63 & $<$ 2.15         &   3.29 $\pm$ 0.92 &  0.98 $\pm$ 0.22 & 10.21 $\pm$ 0.56 \\ 
20 & 26 & 47.62 & 53.26 & 0.75 & $<$ 2.87         &   2.64 $\pm$ 0.81 &  0.69 $\pm$ 0.19 &  8.18 $\pm$ 0.42 \\ 
21 & 27 &  0.79 & 53.25 & 0.87 & $<$ 3.69         &  $<$ 2.48         & $<$ 0.55         &  7.05 $\pm$ 0.40 \\ 
22 & 27 & 13.98 & 53.17 & 0.99 & $<$ 2.25         &  $<$ 3.54         & $<$ 0.55         &  6.31 $\pm$ 0.32 \\ 
23 & 27 & 27.16 & 53.03 & 1.10 & $<$ 3.38         &   2.94 $\pm$ 0.82 & $<$ 0.79         &  5.77 $\pm$ 0.42 \\ 
24 & 27 & 40.34 & 52.80 & 1.22 & $<$ 2.29         &  $<$ 2.30         & $<$ 0.34         &  4.72 $\pm$ 0.30 \\ 
25 & 27 & 53.53 & 52.51 & 1.34 & $<$ 3.60         &  $<$ 2.84         & $<$ 0.48         &  4.08 $\pm$ 0.18 \\ 
26 & 28 &  6.71 & 52.14 & 1.46 & $<$ 2.14         &  $<$ 2.51         & $<$ 0.47         &  3.74 $\pm$ 0.17 \\ 
27 & 28 & 19.89 & 51.71 & 1.58 & $<$ 2.63         &  $<$ 2.52         & $<$ 0.43         &  2.74 $\pm$ 0.16 \\ 
28 & 28 & 33.07 & 51.20 & 1.70 & $<$ 2.96         &  $<$ 2.21         & $<$ 0.36         &  2.48 $\pm$ 0.20 \\ 
29 & 28 & 46.25 & 50.62 & 1.82 & $<$ 2.65         &  $<$ 2.41         & $<$ 0.29         &  2.38 $\pm$ 0.28 \\ 
30 & 28 & 59.44 & 49.97 & 1.94 & $<$ 3.05         &  $<$ 2.52         & $<$ 0.38         &  2.81 $\pm$ 0.19 \\ 
31 & 29 & 12.62 & 49.25 & 2.05 & $<$ 2.34         &  $<$ 2.68         & $<$ 0.26         &  2.51 $\pm$ 0.16 \\ 
32 & 29 & 25.81 & 48.45 & 2.17 & $<$ 2.04         &  $<$ 2.22         & $<$ 0.27         &  2.31 $\pm$ 0.15 \\ \tableline
\enddata
\tablenotetext{a}{16h (J2000)}
\tablenotetext{b}{$-24^\circ 27^\prime$ (J2000)}
\tablenotetext{c}{Distance from HD~147889}
\end{deluxetable}

\clearpage

\begin{deluxetable}{ccccccccc}
\tablecaption{Summary of the H$_2$ line intensities in the $\rho$ Oph region. \label{rOphres_H2}}
\tabletypesize{\scriptsize}
\tablehead{
\colhead{position} & \multicolumn{8}{c}{line intensities ($10^{-8}$\,W\,m$^{-2}$\,sr$^{-1}$)\tablenotemark{b}} \\ 
\colhead{} & \colhead{v=1-0 O(3)} & \colhead{v=2-1 O(3)} & \colhead{v=1-0 O(5)} & \colhead{v=2-1 O(9)} & \colhead{v=0-0 S(3)} & \colhead{v=0-0 S(2)} & \colhead{v=0-0 S(1)} & \colhead{v=0-0 S(0)}\\
\colhead{} & \colhead{2.80\um} & \colhead{2.97\um} & \colhead{3.23\um} & \colhead{4.91\um} & \colhead{9.66\um\tablenotemark{a}} & \colhead{12.3\um} & \colhead{17.0\um} & \colhead{28.2\um}\\
\colhead{} & \colhead{$14^{\prime\prime}\times 20^{\prime\prime}$} & \colhead{$14^{\prime\prime}\times 20^{\prime\prime}$} & \colhead{$14^{\prime\prime}\times 20^{\prime\prime}$} & \colhead{$14^{\prime\prime}\times 20^{\prime\prime}$} & \colhead{$14^{\prime\prime}\times 20^{\prime\prime}$} & \colhead{$14^{\prime\prime}\times 27^{\prime\prime}$} & \colhead{$14^{\prime\prime}\times 27^{\prime\prime}$} & \colhead{$20^{\prime\prime}\times 27^{\prime\prime}$}
}
\startdata
13 & -- & --       &       -- &       -- & $<$ 2.01 &        -- &            -- & $<$ 3.44 \\
14 & -- & $<$ 1.67 & $<$ 1.43 & $<$ 8.76 & $<$ 3.00 & $<$ 23.71 & 3.80$\pm$0.96 & $<$ 5.73 \\
15 & -- & $<$ 1.30 & $<$ 1.44 & $<$ 6.71 & 3.49$\pm$0.61 & $<$ 19.88 & 8.96$\pm$1.00 & $<$ 3.58 \\
16 & -- & $<$ 3.46 & $<$ 2.42 & $<$ 11.49 & 2.79$\pm$0.72 & $<$ 15.78 & 5.53$\pm$0.79 & $<$ 3.50 \\
17 & -- & $<$ 3.28 & $<$ 2.55 & $<$ 9.16 & 5.72$\pm$0.65 & $<$ 23.82 & 12.21$\pm$1.10 & $<$ 4.10 \\
18 & $<$ 3.31 & $<$ 2.34 & $<$ 2.50 & $<$ 15.75 & 8.41$\pm$0.83 & 17.17$\pm$3.14 & 18.15$\pm$1.22 & 4.03$\pm$1.25 \\
19 & -- & -- & $<$ 0.77 & $<$ 5.33 & $<$ 1.98 & -- & 2.52$\pm$0.77 & $<$ 3.09 \\
20 & -- & --       &       -- &       -- & $<$ 1.89 &        -- &            -- & $<$ 2.75 \\
\enddata
\tablenotetext{a}{The upper limit of the 9.66\um~intensities at the observed positions not listed here are $3.0\times 10^{-8}$\,W\,m$^{-2}$\,sr$^{-1}$.}
\tablenotetext{b}{The third row indicates the aperture size of the SWS.  Lines with -- indicate that the observation was not been carried out.}
\end{deluxetable}

\clearpage

\begin{deluxetable}{ccccccccc}
\tablecaption{Summary of the line intensities in the $\sigma$ Sco region. \label{sScores}}
\tablehead{
\colhead{No.}  & \multicolumn{2}{c}{R.A.\tablenotemark{a}} & \multicolumn{2}{c}{DEC.\tablenotemark{b}} & \colhead{$d$\tablenotemark{c}} &
\multicolumn{3}{c}{line intensities ($10^{-8}\,\mbox{W{\,}m}^{-2}\,\mbox{sr}^{-1}$)}\\
\colhead{}& \colhead{m} & \colhead{s} & \colhead{m} & \colhead{s} & \colhead{pc} & \colhead{\oi~63\um} & \colhead{\nii~122\um} & \colhead{\cii~158\um}
}
\startdata
 1 & 19 & 13.73 &  9 &  5.90 & 1.55 &	 1.14 $\pm$ 0.36 & $<$ 0.23         &  2.14 $\pm$ 0.21 \\ 
 2 & 19 & 23.10 & 11 & 13.20 & 1.43 &	 0.82 $\pm$ 0.27 & $<$ 0.18         &  2.68 $\pm$ 0.15 \\ 
 3 & 19 & 32.48 & 13 & 20.50 & 1.30 &	$<$ 1.00         & $<$ 0.21         &  2.82 $\pm$ 0.21 \\ 
 4 & 19 & 41.87 & 15 & 27.70 & 1.18 &	$<$ 1.36         & $<$ 0.27         &  3.74 $\pm$ 0.16 \\ 
 5 & 19 & 51.26 & 17 & 34.90 & 1.05 &	 1.73 $\pm$ 0.29 & $<$ 0.42         &  4.37 $\pm$ 0.20 \\ 
 6 & 20 &  0.65 & 19 & 42.10 & 0.93 &	 1.67 $\pm$ 0.36 & $<$ 0.28         &  4.45 $\pm$ 0.18 \\ 
 7 & 20 & 10.05 & 21 & 49.30 & 0.81 &	 1.14 $\pm$ 0.36 &  0.34 $\pm$ 0.09 &  4.71 $\pm$ 0.23 \\ 
 8 & 20 & 19.46 & 23 & 56.40 & 0.68 &	$<$ 1.23         & $<$ 0.59         &  4.53 $\pm$ 0.19 \\ 
 9 & 20 & 28.87 & 26 &  3.40 & 0.56 &	$<$ 1.07         &  0.44 $\pm$ 0.14 &  4.23 $\pm$ 0.22 \\ 
10 & 20 & 38.29 & 28 & 10.50 & 0.43 &	$<$ 1.32         &  0.40 $\pm$ 0.11 &  4.44 $\pm$ 0.19 \\ 
11 & 20 & 47.71 & 30 & 17.40 & 0.31 &	$<$ 1.04         & $<$ 0.27         &  3.47 $\pm$ 0.13 \\ 
12 & 20 & 57.14 & 32 & 24.40 & 0.19 &	$<$ 1.44         & $<$ 0.25         &  3.29 $\pm$ 0.17 \\ 
13 & 21 &  6.58 & 34 & 31.30 & 0.06 &	$<$ 1.47         &  0.34 $\pm$ 0.08 &  2.46 $\pm$ 0.17 \\ 
14 & 21 & 16.02 & 36 & 38.20 & 0.06 &	$<$ 1.22         & $<$ 0.29         &  2.57 $\pm$ 0.10 \\ 
15 & 21 & 25.46 & 38 & 45.00 & 0.19 &	$<$ 1.06         & $<$ 0.30         &  2.65 $\pm$ 0.10 \\ 
\enddata
\tablenotetext{a}{16h (J2000)}
\tablenotetext{b}{$-25^\circ$ (J2000)}
\tablenotetext{c}{Distance from $\sigma$ Sco}
\end{deluxetable}

\clearpage

\clearpage

\clearpage

\clearpage

\clearpage

\clearpage

\clearpage

\end{document}